# Learning Dynamic MRI Reconstruction with Convolutional Network Assisted Reconstruction Swin Transformer

Di Xu[1], Hengjie Liu[2], Dan Ruan[2] and Ke Sheng[1]

[1] Radiation Oncoloy, University of California, San Francisco
[2] Radiation Oncology, University of California, Los Angeles
{di.xu; ke.sheng}@ucsf.edu

**Abstract.** Dynamic magnetic resonance imaging (DMRI) is an effective imaging tool for diagnosis tasks that require motion tracking of a certain anatomy. To speed up DMRI acquisition, k-space measurements are commonly under-sampled along spatial or spatial-temporal domains. The difficulty of recovering useful information increases with increasing under-sampling ratios. Compress sensing was invented for this purpose and has become the most popular method until deep learning (DL) based DMRI reconstruction methods emerged in the past decade. Nevertheless, existing DL networks are still limited in long-range sequential dependency understanding and computational efficiency and are not fully automated. Considering the success of Transformer's positional embedding and "swin window" self-attention mechanism in the vision community, especially natural video understanding, we hereby propose a novel architecture named Reconstruction Swin Transformer (RST) for 4D MRI. RST inherits the backbone design of the Video Swin Transformer with a novel reconstruction head introduced to restore pixel-wise intensity. A convolution network called SADXNet is used for rapid initialization of 2D MR frames before RST learning to effectively reduce the model complexity, GPU hardware demand, and training time. Experimental results in the cardiac 4D MR dataset further substantiate the superiority of RST, achieving the lowest RMSE of 0.0286±0.0199 and 1-SSIM of 0.0872±0.0783 on 9 times accelerated (9x) validation sequences.

**Keywords:** Transformer, Dynamic MRI, Reconstruction, Deep Learning.

## 1 Introduction

Tracking dynamic processes using time-resolved magnetic resonance imaging (MRI) can reveal anatomical and physiological anomalies that evade detection based on static images [1]. Theoretically, a dynamic target can be acquired frame-wise under the assumption that motion in each frame at the time of acquisition is within the chosen pixel size. However, the assumption of which requires extremely high spatial-temporal (ST) resolutions and thus is often violated due to slow data acquisition speed [2].

To speed up the process of dynamic MRI (DMRI) acquisition, a wide range of approaches have been proposed. Advanced fast imaging sequences [3] and modern gradient system allow efficient data sampling in k-space. Supported by massive coil arrays [4], parallel imaging [5, 6] has increased scanning speed by 2-4 folds in clinical practice. Followed by that, further acceleration has been achieved through reducing temporal [7, 8] or spatiotemporal [9, 10] redundancy in dynamic images, enabled through the design of special k-space under-sampling trajectories (e.g. variable density Cartesian, radial, spiral and etc.). Those methods heavily rely on effective reconstruction to recover artifact-controlled images [11].

Among them, Compressed sensing (CS) [12, 13] has demonstrated its potential in MRI acceleration. CS methods were developed by solving a constrained optimization problem, including priors as regularization to stabilize the solution [2, 14–16]. Pertinent to the dynamic MRI problem, for cohesion in $T$ axis, several studies combined motion estimation/compensation (ME/MC) in a unified reconstruction framework for more effective acceleration [1, 17]. More recently, Zhao et al. proposed to integrate an intensity-based optical flow prior into the optimization so that motion is compensated through the optical flow estimated motion field [1]. Many studies show that properly implemented CS methods can reconstruct artifacts-free images with k-space data not meeting the Nyquist condition. On the other hand, CS methods share two common weaknesses. First, being iterative methods, they tend to be slow. Second, the results depend on heuristic hyperparameter (HP) tuning. Because of these limitations, CS is often incompatible with the clinical needs for an efficient, automated, and generalizable solution [18].

Recently, research has pivoted to a new direction using deep learning (DL) models for DMRI restoration. In specific, Schlemper et al. extended a 2D convolutional neural network (CNN) based MR reconstruction framework into 3D and then introduced a data-sharing layer derived from K nearest neighbor (KNN) to regularize the temporal correlation among frames [19]. Their CNN model can recover information on individual 2D frames, but the KNN data-sharing layer is insufficient to robustly and efficiently estimate the dynamic relationship among frames. Nonparametric algorithms, including KNN, are more likely to suffer from overfitting, sensitive to HP tuning, and are computationally expensive [20]. To better conduct ME/MC in DL-based DMRI reconstruction, researchers have started investigating the potentiality of recurrent neural networks (RNNs) for temporal pattern modeling. For DMRI reconstruction, Qin et al. deployed the conventional iterative reconstruction scheme with an RNN, with recurrent hidden connections capturing the iterative reconstruction and bidirectional hidden connections characterizing spatial-temporal dependencies [21]. Huang et al. introduced a motion-guided network using ConvGRU to reconstruct initial images $I$ in ST space and then U-FlowNet to perform MC to $I$ through learning optical flow [22]. Yet, RNN models suffer from practical failure in learning long dependencies and bi-directional information that are undesirable in the modern concept of large-scale parallel training. Moreover, RNNs perform sequential learning, assuming each state to be dependent only on the previously seen states, which mandates the sequences to be processed frame by frame and limits encoding of a specific frame to the next few evolutions.

Transformer was introduced to overcome these drawbacks in machine translation, aiming to avoid recursion, allow parallel computation, and reduce performance drops due to long sequential dependencies [23]. Subsequently, the vision community has seen a modeling shift from CNNs to Transformers which have attained top accuracy in numerous video processing benchmarks [24]. Apart from the preeminence in positional embedding, Video Swin Transformer (VST), one of the frameworks benefitting video understanding, uses a novel hierarchical shifted window mechanism, which performs efficient self-attention in the ST domain using non-overlapping local windows and cross-window connection [25]. Since DMRI can be considered a form of video data, we propose extending VST to the DMRI reconstruction task.

To this end, we developed a novel Reconstruction Swin Transformer (RST) network for four-dimensional (4D) MR recovery. To accelerate the training speed and reduce the complexity of RST, we employed a CNN-based 2D reconstruction network named SADXNet[26], for rapid recovery of MR frames prior to feeding into RST. To our knowledge, this is the first work that introduces self-attention and positional embedding to DMRI reconstruction. Experiments on a cardiac dataset have demonstrated the superiority of our proposed architecture over several baseline approaches.

## 2 Materials and Methods

### 2.1 SADXNet Assisted Reconstruction Swin Transformer

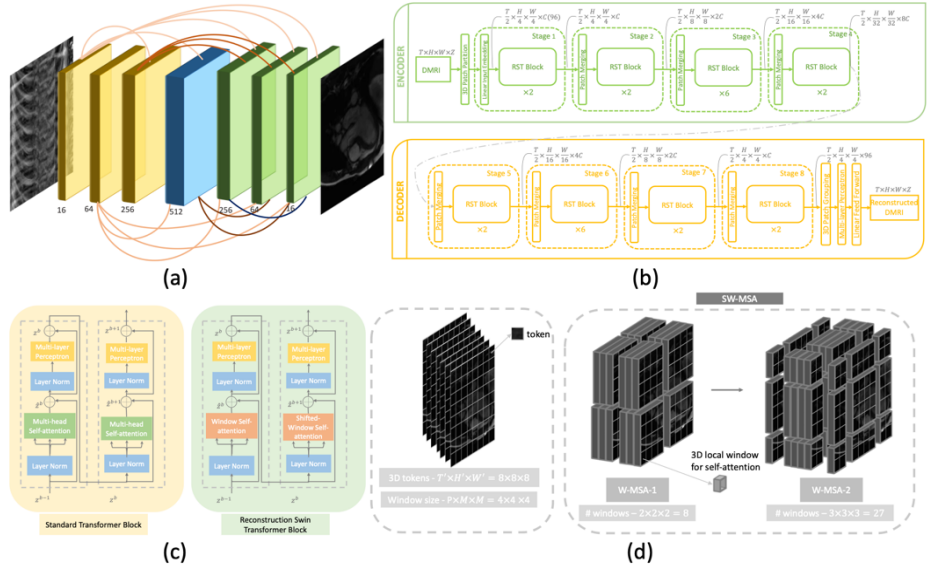

**Fig. 1.** (a) SADXNet architecture. (b) RST-T structure. (c) Comparison of RST and standard Transformer blocks. (d) An instance of the SW-MSA layer.

Our framework consists of two stages. 1) Initial 2D frame-wise reconstruction using SADXNet [26]. 2) 4D tensor learning in the ST domain with RST. Theoretically, RST can yield comparable performance without SADXNet initialization conditioned upon unlimited network parameters, training time, and GPU capacity. Nevertheless, numerous studies showed that training on high-resolution 4D tensors requires significantly more parameters to fully generalize the feature representation, which takes overly long training time to converge, deviates from the clinical aim of sub-second prediction, and is beyond most of the current commercial GPU memories [27, 28]. Thus, we proposed to use SADXNet for spatial information recovery, followed by a comparable lighter-weighted RST for collective spatial-temporal data reconstruction. We elaborate on the structures of SADXNet and RST as follows.

**SADXNet.** SADXNet was originally introduced as an image-denoising network in [26] for chest x-ray rib suppression. Noting that the tasks of image denoising and reconstruction fundamentally work on pixel-level recovery, we applied SADXNet to reconstruct MR frames. SADXNet is designed to be fully convolutional and densely connected, as shown in Fig. 1 (a). Firstly, it has seven densely connected layers designed to be channel-wise concatenated, where feature maps produced from a specific layer and all its preceding layers are concatenated in the channel dimension and fed into all its subsequent layers. Secondly, three consecutive operations are conducted for each convolution layer, including batch normalization (BN), rectified linear unit (ReLU), and convolution with a filter size of 3×3. Thirdly, no up- or down-sampling is implemented to preserve the spatial dimension of feature maps through the forward flow. Fourthly, the convolution kernels in the seven layers are organized with an increased-to-decreased number of channels. Lastly, the cost function for SADXNet is formulated as a combination of negative peak signal-to-noise ratio (PSNR), muti-scale structure similarity index measure (MS-SSIM), and $L_1$ deviation measurement as shown in Equation (1-4) [29].

$$L = -\alpha \cdot L_{PSNR} + (1-\alpha) \cdot [\beta \cdot (1 - L_{MS\text{-}SSIM}) + (1-\beta) \cdot L_1] \quad (1)$$

$$L_{PSNR} = log_{10}\left(\frac{MAX_X^2}{1/mn \cdot \sum_{j=0}^{m-1}\sum_{i=0}^{n-1}[x_{ij}-y_{ij}]^2}\right) \quad (2)$$

$$L_{MS-SSIM} = \frac{(2\mu_x\mu_y+c_1)}{(\mu_x^2+\mu_y^2+c_1)} \cdot \prod_{j=1}^{M}\frac{(2\sigma_{x_j y_j}+c_2)}{(\sigma_{x_j}^2+\sigma_{y_j}^2+c_2)} \quad (3)$$

$$L_1 = \frac{1}{mn} \cdot \sum_{j=0}^{m-1}\sum_{i=0}^{n-1}||x_{ij}-y_{ij}||_1 \quad (4)$$

Where $\alpha$ and $\beta$, set as 0.5 and 0.5 respectively, are tunable HPs, $x$, and $y$ are input and supervision, $MAX_X$ is the maximum possible input value, $S$ is the dynamic range of the pixel values, $(k_1, k_2)$ are two constants, and $||\cdot||_1$ is the $l_1$ norm.

**Reconstruction Swin Transformer.** RST strictly follows the hierarchical design of the VST backbone. We introduce a novel reconstruction head in the RST decoder stage to fulfill the goal of pixel-wise information recovery. As a variant of VST, RST also has four architectures with distinct model complexities. 1) RST-T: $C = 96$, block numbers = {2,2,6,2, 2,6,2,2}; 2) RST-S: $C = 96$, block numbers = {2,2,18,2, 2,18,2,2}; 3) RST-B: $C = 128$, block numbers = {2,2,18,2, 2,18,2,2}; 4) RST-L: $C = 192$, block numbers = {2,2,18,2, 2,18,2,2}, where $C$ represents the channel number of the first RST block [25].

We illustrate the architecture of RST-T in Fig. 1(b). First, the backbone of RST consists of four stages and performs 2 × down-sampling solely in each block's spatial dimension of the patch merging layer [25]. The RST blocks within the backbone and head are symmetrically parameterized, except that all the patch merging layers in the reconstruction head perform 2 × spatial up-sampling. Second, for an input 4D tensor of dimension $T \times H \times W \times Z$, we define each patch of size $2 \times 4 \times 4 \times Z$ as a 3D token, where the $Z$-axis of DMRI is considered as the color channel in natural images. Thus, after processing through the initial 3D patch partition layer, we obtain $\frac{T}{2} \times \frac{H}{2} \times \frac{W}{2} \times Z$ 3D tokens. Third, a linear embedding layer is applied right after 3D patch partition to project the channel dimension from $Z$ to pre-specified $C$ with respect to the destinated RST variant, which is 96 in RST-T.

The critical component of the "swin window" mechanism lies in the RST block, which replaces the multi-head self-attention (MSA) layer in the block of the standard Transformer with 3D shifted window-based MSA (SW-MSA) and keeps the rest subcomponents unaltered (see Fig. 1 (c)) [25]. The design of SW-MSA introduces a locality inductive bias in its self-attention module to accommodate additional temporal information intake. SW-MSA scheme is realized with two consecutive W-MSA modules. In the first W-MSA, given an initial input composed of $T' \times H' \times W'$ 3D tokens and a window of size $P \times M \times M$, windows are arranged in a manner that can non-overlappingly partition the input tokens. Next, a 3D-shifted window is introduced immediately following the W-MSA layer to establish cross-connections among various windows. Given the $\frac{T'}{P} \times \frac{H'}{M} \times \frac{W'}{M}$ windows obtained from the first W-MSA having size of $P \times M \times M$, the partition configuration of the second W-MSA will shift the windows along the ST axis by $\frac{P}{2} \times \frac{M}{2} \times \frac{M}{2}$ tokens from the partition coordinates in the first W-MSA. In Fig. 1 (d), given $8 \times 8 \times 8$ tokens and a window of size $4 \times 4 \times 4$, W-MSA-1 will partition the tokens into $\frac{T'}{P} \times \frac{H'}{M} \times \frac{W'}{M} = 2 \times 2 \times 2 = 8$ windows. Next, the windows are shifted by $\frac{P}{2} \times \frac{M}{2} \times \frac{M}{2} = 2 \times 2 \times 2$ tokens, resulting in $3 \times 3 \times 3 = 27$ windows in W-MSA-2.

The cost function of RST is designed to be identical to that of SADXNet in Equation (1-4), which is a combination of PSNR, MS-SSIM, and $L_1$ deviation.

**Experimental Setup and Baseline Algorithms.** Both SADXNet and RST were implemented in Pytorch, and the training was performed on a GPU cluster with $4 \times RTX\ A6000$. For the training of SADXNet, data augmentation, including random rotation, resizing, cropping, and Gaussian blurring, was implemented. Adam optimizer with an initial learning rate (LR) of 0.001, an epoch of 1000, and a batch size of 2×4 was applied during learning. RST-S without SADXNet initialization was trained, and vice versa for RST-T. Data augmentation was implemented for all RST training, including random cropping, resizing, and flipping. Adam optimizer with an initial LR of 0.001 and batch size of 1×4 was applied.

We compared with four state-of-the-art approaches, including k-t SLR [2], k-t FOCUSS+ME/MC [30], DC-CNN (3D) [19], and MODRN (e2e) [22]. The first two are conventional CS-based methods, where k-t FOCUSS+ME/MC includes ME/MC procedures. The latter two are DL methods where DC-CNN (3D) explores ST information using 3D convolution and MODRN (e2e) includes ME/MC into their RNNs. Three quantitative metrics are used for evaluation: rooted mean squared error (RMSE), PSNR, and SSIM.

## 2.2 Data Cohort

The open-access cardiovascular Magnetic resonance (OCMR) dataset provides multi-coil Cartesian-sampled k-space data from 53 fully sampled and 212 prospectively under-sampled cardiac cine series with both short-axis (SAX) and long-axis (LAX) views. The fully sampled scans have average frames of 18, and the under-sampled scans each have 65 frames. The dataset was collected on three Siemens MAGNETOM scanners - Prisma (3T), Avanto (1.5T), and Sola (1.5T) – with a 34-channel receiver coil array, using bSSFP sequence with FOV=800 × 300 mm, acquisition matrix 384 × 144, slice thickness = 8 mm, and TR/TE=38.4/1.05 ms. The fully sampled data was collected with ECG gating and breath-holding. The under-sampled data with an acceleration rate (AR) of 9 × was collected under the free-breathing condition in real-time. More details can be found in [31]. We prepared 3 subsets, including training, validation, and test sets to assess our framework. The training and validation sets were split from the 53 fully sampled sequences with balanced long/short-axis views and scanner types at the ratio of training/validation = 40/13, whereas the 212 prospectively under-sampled data was used for testing. Variable-density incoherent spatial-temporal acquisition (VISTA) [32] was used to retrospectively under-sample the training/validation sequences.

## 3 Results

For the OCMR dataset, performance comparison was made among SADXNet initialized RST-T, solely RST-S, and the selected benchmark algorithms - k-t-SLR, k-t-FOCUSS-ME/MC, DC-CNN, and MODRN (e2e). We respectively trained SADXNet+RST-T and RST-S for around 2000 and 8000 epochs till observing training convergence.

Quantitative results and visualization of the validation set are reported in Table 1 and Fig.2 (a). Solely visualization on the prospective test set of OCMR is reported in Fig. 2 (b-c) since the GTs of the test set are not available. Overall, statistical results in Table 1 and visual outcomes in Fig. 2 (a-b) consistently show that DL models, including DC-CNN, MODRN (e2e), RST-S, and RST-T, significantly outperform CS methods (k-t-SLR and k-t-FOCUSS+ME/MC). In addition to the promising performance illustrated by the quantitative metrics in Table 1, we can also observe that the DL predictions in Fig. 2 have fewer residuals and streaking artifacts, sharper edge delineation, detailed morphology (arrows pointed), and better concentration of optical flow estimated motion on the cardiac region of interest (ROI) than the ones made by CS methods. Those imply that DL methods have better dynamic motion understanding than the selected CS baselines. Meanwhile, all the DL approaches take considerably less time to reconstruct incoming dynamic imaging than CS algorithms, as seen in Table 1.

Within the two CS-based methods, the predictions from k-t-FOCUSS+ME/ME are marginally better than those from k-t-SLR. Among all listed neural networks, the performance ranking can be summarized as SADXNet initialized RST-T > RST-S > MODRN (e2e) > DC-CNN with SADXNet+RST-T, RST-S, and DC-CNN able to make a sub-second prediction. Additionally, though trained significantly longer for 8000 epochs, RST-S still has difficulty capturing finer morphologies and subtle dynamic movements compared with SADXNet + RST-T, which further substantiates the importance of CNN 2D frame initialization prior to RST training in practice. Finally, we observe that predicted frames from SADXNet+RST are consistent and steady along the T axis in Fig. 2 (c), where the robustness of our proposed framework is demonstrated.

**Table 1.** Results of OCMR validation sets. Mean values and standard deviation are reported. Results from the best performer are bolded, whereas those from the underperformer are underlined. The time reported is the averaged prediction time for each validation sequence.

| Algorithm | RMSE↓ | PSNR↑ | 1-SSIM↓ | Time (s) | Device |
|---|---|---|---|---|---|
| k-t SLR | 0.132±0.105 | 19.792±5.274 | 0.327±0.276 | 200.731 | CPU |
| k-t FOCUSS + MC/ME | 0.0873±0.0572 | 21.245±6.143 | 0.201±0.243 | 256.874 | CPU |
| DC-CNN | 0.0552±0.0431 | 25.796±4.342 | 0.145±0.0872 | 0.543 | GPU |
| MODRN (e2e) | 0.0392±0.0311 | 29.341±3.776 | 0.121±0.0788 | 5.736 | GPU |
| RST-S | 0.0323±0.0353 | 30.868±3.104 | 0.114±0.0996 | **0.337** | GPU |
| SADXNet + RST-T | **0.0286±0.0199** | **33.587±2.991** | **0.0872±0.0783** | 0.681 | GPU |

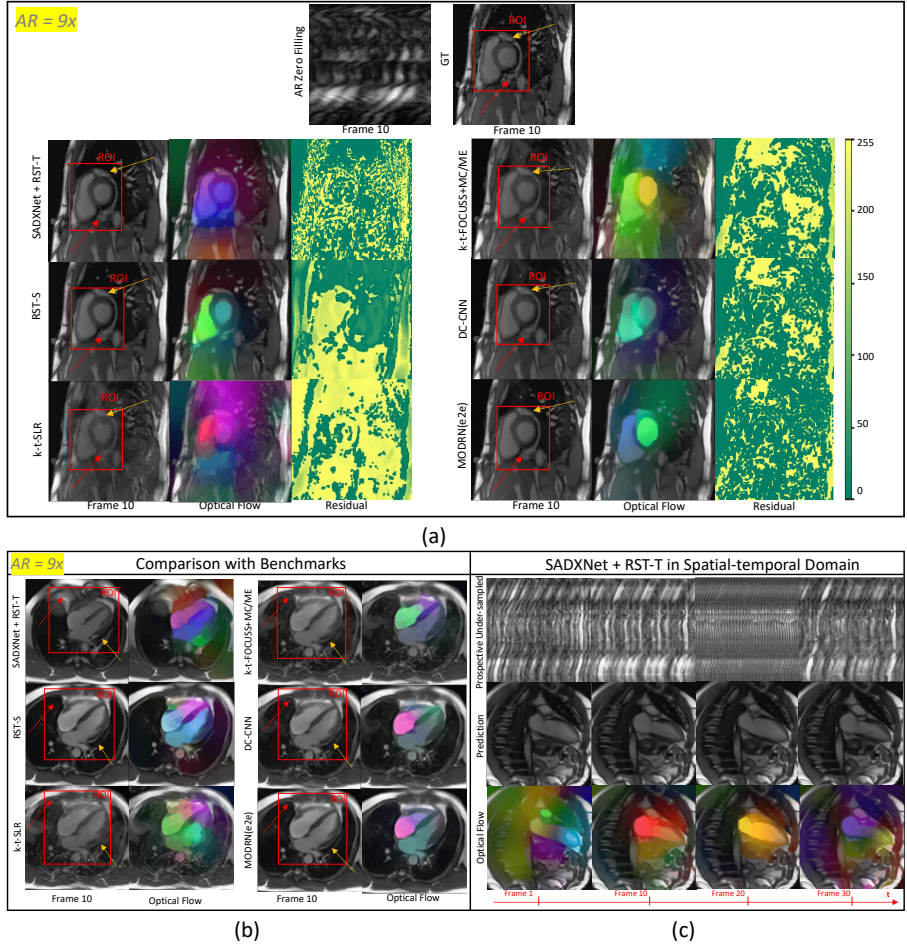

**Fig. 2.** (a) Visualization of reconstruction results on a SAX sequence in the validation set. (b) The reconstruction results on a LAX four-chamber-view sequence in the test set (AR=9x). In both (a) and (b), outcomes were visualized having each panel consisting of predictions with and without optical flow overlayed on the 10[th] frame. ROIs are bounded out in red with arrows pointing out the comparison of artifact and morphology. (c) Visualization of frames 1-30 from a LAX two-chamber view sequence predicted by SADXNet+RST-T.

## 4 Discussion

The current study presents a novel RST architecture adapted from VST for DMRI reconstruction. RST inherits the backbone of VST and introduces a novel reconstruction head to fulfill the demand for pixel-wise intensity prediction. Compared with existing DL-based reconstruction approaches [19, 22], RST markedly improved efficiency and efficacy. The efficiency stems from the positional embedding scheme of the Transformer, which enables fully distributable processing in the spatial and temporal domains, resulting in enormously reduced training and prediction time. The efficacy is boosted via SW-MSA, which conducts self-attention learning in both spatial and temporal dimensions. Additionally, the SADXNet for the initial restoration of 2D MR frames substantially reduced the training burden, model parameters, and GPU momory footprint. RST-based algorithms outperformed the comparison state-of-the-art CS and DL reconstruction methods by a large margin in the experiments reconstructing the unsampled OCMR dataset. Compared to training RST-S on raw under-sampled images, SADXNet-initiated RST-T delivered an additional performance boost in statistical and visual outcomes while significantly reducing training and model complexity. The results thus support the importance of 2D frame initialization prior to official training from the 4D domain.

Our method is not without room for improvement. SADXNet was introduced to reduce the computation demand of RST, but the current framework still requires substantial GPU memory restricted to higher-end non-consumer grade hardware. As SADXNet demonstrates the promise of RST parameter reduction with spatial domain initialization, we will explore temporal domain initialization before conducting joint 4D ST dimensional learning in future studies.

## 5      Conclusion

A dynamic MR reconstruction framework, SADXNet-assisted RST, is proposed with accuracy enabled through the "swin window" mechanism and efficiency guaranteed from Transformer positional embedding and CNN spatial initialization. Validation from the OCMR dataset substantiates the superior performance of SADXNet+RST to state-of-the-art CS and DL methods.